\documentclass[journal]{IEEEtran}
\usepackage{graphicx}

\begin{document}
\title{Research in Geant4 electromagnetic physics design, and its effects 
on computational performance and quality assurance}
%
%

\author{M. Augelli, M. Begalli, S. Hauf, C. H. Kim,
        M. Kuster, M. G. Pia, P. Queiroz Filho, L. Quintieri, P. Saracco, 
        D. Souza Santos, G. Weidenspointner, A. Zoglauer
\thanks{Manuscript received November 17, 2009.} 
\thanks{Mauro Augelli is with the Centre d'Etudes Spatiales (CNES), Toulouse,
	France.}
\thanks{Marcia Begalli is with State University of Rio de Janeiro, Brazil.}
\thanks{Steffen Hauf and Markus Kuster are with University of Technology,
	Darmstadt, Germany.}
\thanks{Chan Hyeong Kim is with Hanyang University, Seoul, Korea.}
\thanks{Maria Grazia Pia and Paolo Saracco are with INFN Sezione di Genova, 
	Via Dodecaneso 33, 16146 Genova, Italy (e-mail:
	MariaGrazia.Pia@ge.infn.it, Paolo.Saracco@ge.infn.it).}
\thanks{Pedro Queiroz Filho and Denison Souza Santos are with Institute 
	for Radiation Protection and Dosimetry, IRD, Rio de Janeiro, Brazil.}
\thanks{Lina Quintieri is with INFN Laboratori Nazionali di Frascati, 
	Frascati, Italy.}
\thanks{Georg Weidenspointner is with the Max-Planck-Institut f\"ur
	extraterrestrische Physik, Garching, Germany, and
	with the MPI Halbleiterlabor.}
\thanks{Andreas Zoglauer is with the Space Sciences Laboratory,
	University of California at Berkeley, Berkley, CA, USA.}%
}

\maketitle
\pagestyle{empty}
\thispagestyle{empty}

\begin{abstract}
The Geant4 toolkit offers a rich variety of electromagnetic physics
models; so far the evaluation of this Geant4 domain has been mostly
focused on its physics functionality, while the features of its design
and their impact on simulation accuracy, computational performance and
facilities for verification and validation have not been the object of
comparable attention yet, despite the critical role they play in many
experimental applications.
A new project is in progress to study the application of new design
concepts and software techniques in Geant4 electromagnetic physics,
and to evaluate how they can improve on the current simulation
capabilities.
The application of a policy-based class design is investigated as a
means to achieve the objective of granular decomposition of processes;
this design technique offers various advantages in terms of
flexibility of configuration and computational performance. The
current Geant4 physics models have been re-implemented according to the
new design as a pilot project.
The main features of the new design and first results of performance
improvement and testing simplification are presented; they are
relevant to many Geant4 applications, where computational speed and
the containment of resources invested in simulation production and
quality assurance play a critical role.

\end{abstract}


\section{Introduction}
\IEEEPARstart{G}{eant4} \cite{g4nim,g4tns} is an object oriented toolkit for the
simulation of particle interactions with matter. It provides advanced
functionality for all the domains typical of detector simulation:
geometry and material modelling, description of particle properties,
physics processes, tracking, event and run management, user interface
and visualisation.

Geant4 is nowadays a mature Monte Carlo system and is
used in many, multi-disciplinary experimental applications;
its rich collection of physics processes and models, extending
over a wide energy range, has played a key role in satisfying the
needs of a large variety of experimental developments.

Nevertheless, 
new experimental requirements have emerged in the recent years, which
challenge the conventional scope of major Monte Carlo transport codes
like Geant4. Research in nanodosimetry, nanotechnology-based
detectors, radiation effects on components in space and at high
luminosity colliders, nuclear power, plasma physics etc. have
shown the need of new methodological approaches to radiation transport
simulation along with new physics functionality in Geant4.
A common requirement in all such research domains is the ability 
to change the scale at which the problem is treated
in a complex simulation environment. 
Significant technological developments both in software and
computing hardware have also occurred since the RD44 \cite{rd44} phase, 
which defined Geant4 design. 
New software techniques are available nowadays, that were not yet
established at the time when Geant4 was designed.

A R\&D project, named NANO5,
has been recently launched \cite{mc2009,nano5_nss2009} to address fundamental
methods in radiation transport simulation; it explores possible
solutions to cope with the new experimental requirements 
and evaluates whether and how they can be
supported by Geant4 kernel design.
The main focus of the project lies in the simulation at different
scales in the same experimental environment: this objective is
associated with the research of transport methods across the current
boundaries of condensed-random-walk and discrete transport schemes.

This study requires electromagnetic physics processes, and related
physics objects, to be lightweight and easily configurable: one of the
main issues to be addressed in the project is indeed the capability of
objects to adapt dynamically to the environment. 
For this purpose a
pilot project has been set up to evaluate the current design of Geant4
electromagnetic package in view of the foreseen extension of capabilities:
it
investigates design techniques suitable to better support fine-grained
physics customization and mutability in response to the environment.

The project adopts a software process
model based on the Unified Process \cite{up} framework.
The software developments are motivated by concrete
experimental applications, and significant effort is invested in the
software design: these features of the project are well served by the
Unified Process, which is use case driven and architecture-centric.
The adopted software process framework involves an iterative and
incremental life-cycle.

\section{Generic programming techniques in physics simulation design}

Metaprogramming has emerged in the last few years as a powerful design
technique. In C++ the template mechanism provides naturally a rich
facility for metaprogramming; Boost libraries \cite{boost} are
nowadays easily available to support generic programming development. 

Metaprogramming presents several interesting advantages,
which propose it as a worthy candidate for physics simulation design.
This technique has not been exploited in Geant4 core yet, 
the partial
support of templates available in C++ compilers at the time of the RD44 phase
was a limiting factor in the
exploitation of templates in Geant4 architectural design at that stage.  
A preliminary investigation of generic programming techniques in a
multi-platform simulation context has been carried out
through the application of a policy-based class
design \cite{dna}; this prototype was limited to a small physics sub-domain.

An advantage of generic programming techniques 
over conventional object oriented programming is the
potential for performance improvement. 
Physics modelling specialization would profit of the shift from
dynamic to static polymorphism, which binds it at compile time rather
than runtime, thus resulting in intrinsically faster programs.

Design techniques intrinsically capable of performance gains are
relevant to computationally intensive simulation domains, like
calorimetry and microdosimetry; in general, the large scale simulation
productions required by HEP experiments would profit of
opportunities for improved physics performance. 

A side product of the use of generic programming techniques in Geant4
design is the improved transparency of physics models: the technology
intrinsically achieves their exposure at a fine-grained level.
This feature greatly facilitates the validation of
the code at microscopic level and the flexible configuration of
physics processes in multiple combinations.

Customization and extensibility through the provision of user-specific
(or experiment-specific) functionality in the simulation are also
facilitated by this technique: in fact, metaprogramming allows the user to write more
expressive code, that more closely corresponds to the mental model of
the problem domain, like the configuration of physics
modeling options in experimental applications.

As a side benefit, a design based on this technique
would naturally overcome all the current issues about duplicated or
competing functionality in different Geant4 physics packages.

It is worth reminding the reader
that, since dynamic and static polymorphism coexist in C++, the
adoption of generic programming techniques would not force Geant4
developers and users to replace object oriented methods entirely: a
clever design can exploit generic and object oriented programming
techniques in the same software environment according to the
characteristics of the problem domain.

Generic programming appears a promising candidate technique to support
the design of the discrete simulation sector in an efficient,
transparent and easily customizable way; the agile design achievable
with such techniques would greatly facilitate the kernel evolution to
accommodate both condensed-random-walk and discrete schemes.

\section{Prototype design}

The R\&D project currently elaborates a conceptual scheme for condensed
and discrete simulation approaches to co-work in the same environment,
and a software design capable of supporting it. This requirement
implies the introduction of a new concept in the simulation: mutable
physics entities (process, model or other physics-aware object), whose
state and behavior depend on the environment and may evolve as an
effect of it. Such a new concept requires rethinking how the Geant4 kernel
handles the interaction between tracking and processes, and represents
a design challenge in a Monte Carlo software system.

The introduction of the concept of mutability in physics-related objects
requires the identification of their stable and mutable states and behaviour,
and their fine-grained decomposition into parts capable of evolving, or 
remaining unchanged.

The first step along this path involves the re-design of the current Geant4
electromagnetic processes \cite{lowe_chep,lowe_nss}, \cite{standard}..
Processes are decomposed down to fine granularity, and objects
responsible of well-identified functionality are created.
The fine-grained decomposition of processes is propaedeutic to handling 
their stable and mutable components independently.

The application of a policy-based class design \cite{alexandrescu} is currently
investigated as a means to achieve the objective of granular
decomposition of processes. This design technique offers various
advantages in terms of flexibility of configuration and computational
performance; however, its suitability to large scale physics
simulation and its capability to model the evolution associated with
mutable physics entities have not been fully demonstrated yet.

For this purpose, a pilot project is currently in progress in the
domain of photon interactions (Compton and Rayleigh scattering,
photoelectric effect and photon conversion): the current Geant4
physics models are re-implemented in terms of the new design, thus
allowing performance measurements as well as first-hand evaluations of
the capabilities and drawbacks of the policy-based design.

The design prototype has adopted a minimalist approach. A generic
process acts as a host class, which is deprived of intrinsic physics
functionality. Physics behavior is acquired through policy classes,
respectively responsible for cross section and final state
generation. A UML (Unified Modelling Language) \cite{uml} class diagram
illustrates the main features of the design in Figure 1.

\begin{figure*}
\begin{center}
\includegraphics[angle=0,width=16cm]{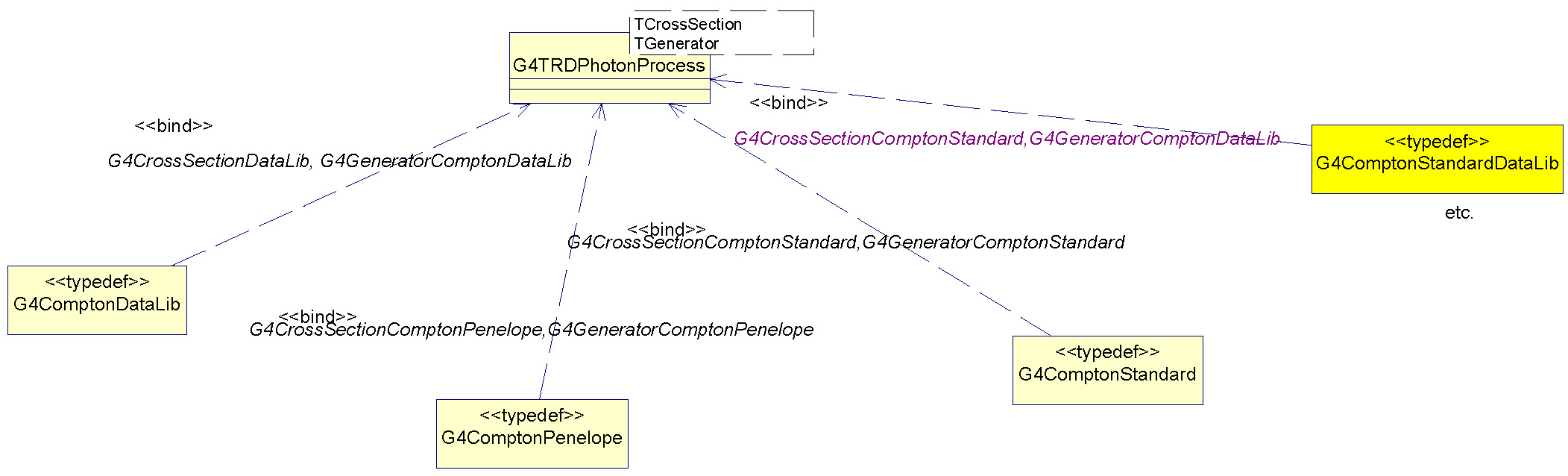} 
\caption{Main features of the policy-based design prototype,
illustrated for Compton scattering.}
\end{center}
\label{fig_em_typedef}
\end{figure*}

The design of the policy classes reimplementing 
the existing physics functionality is currently focused on the 
adoption of well-established object oriented programming practices.
Basic software requirements, like encapsulation and sharply identified
object responsibilities, are enforced throughout the design.
The exploration of more sophisticated software solutions, as well as possible
improvements or extensions to the current physics functionality, are not 
a priority in the first development cycle now in progress, but 
greater attention could be devoted to these aspects in future development
cycles, once the basic design issues have been addressed in a working
prototype.

\section{Preliminary results}

Fully functional processes for photon interactions can be configured at 
the present stage in the new
design by assembling fine-grained policy classes into a generic host class.
A few metrics have been collected to evaluate quantitatively 
the possible benefits or drawbacks of the candidate software technology.

Preliminary performance measurements in a few simple physics test
cases concerning photon interactions indicate a gain on the order of
30\% with respect to equivalent physics implementations in the current
Geant4 design scheme; however, it should be stressed that no effort
has been invested yet into optimizing the new design prototype, nor
the code implementation.

The testing of basic physics components of the simulation is also
greatly facilitated with respect to the current Geant4 version: since
in the new design scheme they are
associated with low level objects like policy classes, they can be
verified and validated independently. 
This agility represents an improvement over the current design
scheme, where a full-scale Geant4-based application is necessary to study
even low-level physics entities of the simulation, like atomic cross
sections or features of the final state models.

A quantitative appraisal of this improvement was performed by
comparing the effort needed to compare Geant4 cross sections of photon
interactions against NIST reference data: this comparison, together
with similar tests concerning electron stopping powers and ranges,
documented in
\cite{g4nist}, requires a Geant4-based application consisting of 
approximately 4000 lines of code in a conventional scheme of Geant4
electromagnetic physics design, whereas the test code required for the
comparison of photon cross sections in the proposed policy-based class
design amounts to less than 100 lines.
Similarly, the simulation production associated with the results described in
\cite{g4nist} required a significant investment of resources (a dedicated test
application developer, a simulation production responsible and the computing
resources of a PC farm), whereas the tests associated with the policy-based
design can be executed in terms of minutes of human time on a laptop
computer.

The inter-comparisons of Geant4 physics models is also greatly facilitated; 
an example is shown in Figures 2 to 4, concerning Compton scattering.

\begin{figure}
\includegraphics[angle=0,width=9cm]{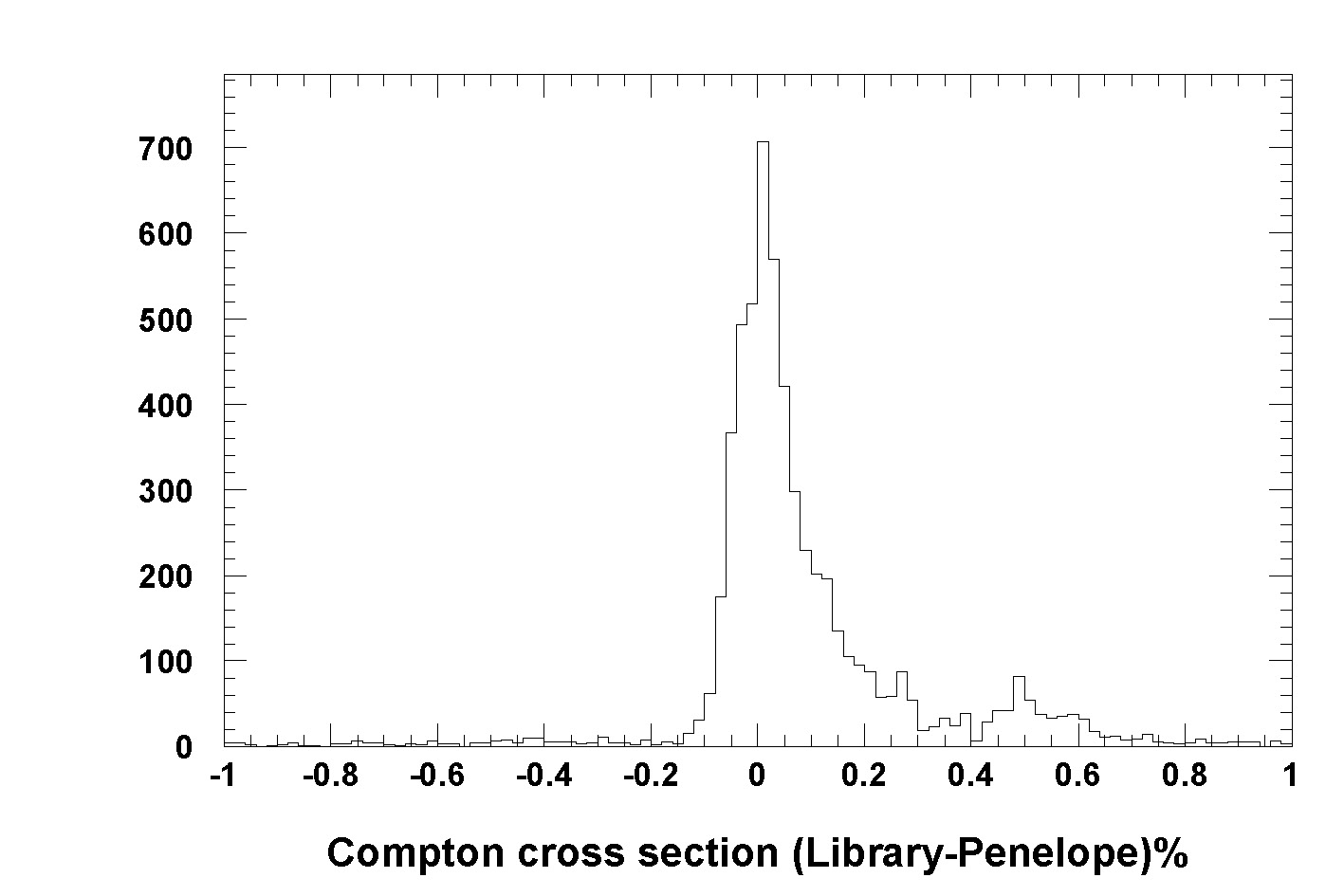} 
\caption{Percent difference between Library-based \cite{lowe_chep,lowe_nss}
and Penelope-like \cite{penelope}
Compton scattering
cross sections over the energy range 1 keV to 100 GeV; the cross sections
are calculated through policy-based classes implementing the same 
functionality as the physics processes and models released in Geant4 9.1.}
\label{fig_cross_libpen}
\end{figure}

\begin{figure}
\includegraphics[angle=0,width=9cm]{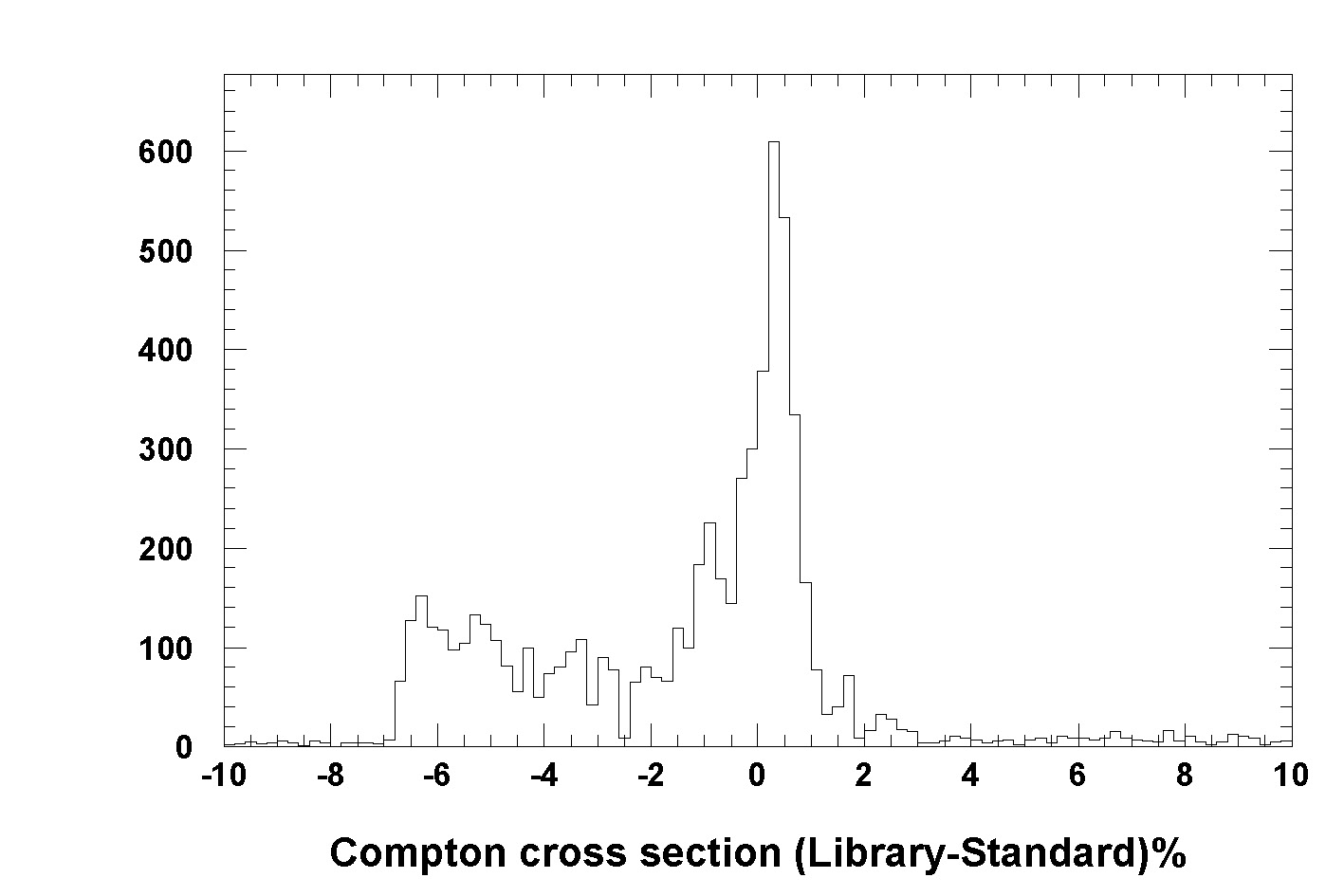} 
\caption{Percent difference between Library-based \cite{lowe_chep,lowe_nss}
and Standard \cite{standard}
Compton scattering
cross sections over the energy range 1 keV to 100 GeV; the cross sections
are calculated through policy-based classes implementing the same 
functionality as the physics processes and models released in Geant4 9.1.}
\label{fig_cross_libstd}
\end{figure}

\begin{figure}
\includegraphics[angle=0,width=9cm]{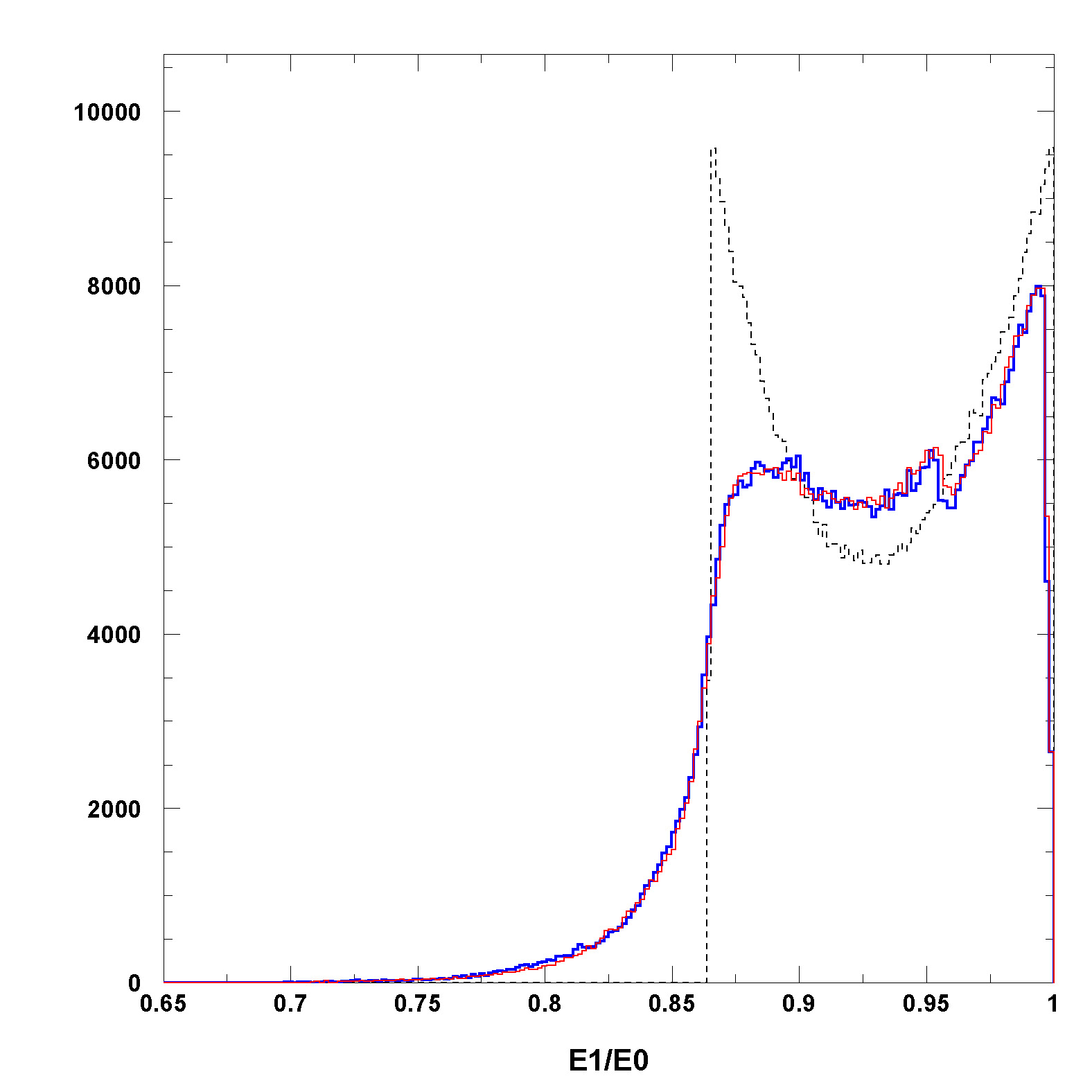} 
\caption{Differential cross section of Compton
scattering as a function of the ratio of the scattered photon's energy
over the incident one's, for 40 keV photons impinging onto a silicon
target; the distributions are calculated by policy-based classes
implementing the same functionality as Library-based  
\cite{lowe_chep,lowe_nss} (blue),
Penelope-like (red) \cite{penelope}
 and Standard (black) \cite{standard}
physics models released in Geant4 9.1.}
\label{fig_doppler}
\end{figure}

\section{Conclusion and outlook}

A R\&D project is in progress to address the capability of handling
multi-scale use cases in the same simulation environment associated
with Geant4: this requirement involves the capability of handling
physics processes according to different transport schemes.
A propaedeutic investigation is in progress to evaluate design
techniques, like generic programming, capable of supporting the main
design goals of the project.

A pilot project concerns the re-design of Geant4 photon interactions,
to evaluate conceptual methods and design techniques suitable to
larger scale application. 

Preliminary results on the application of policy-based design techniques
indicate that
significant improvement in the flexibility of the physics design is
achieved along with a non-negligible improvement in execution time and
facilitated verification and validation testing.
At the present time no adverse effect memory
consumption has been observed yet in association with the prototype
design.

\section*{Acknowledgment}
The authors thank Sergio Bertolucci, 
Thomas Evans,
Elisabetta Gargioni,
Simone Giani, 
Vladimir Grichine,
Bernd Grosswendt,
Andreas Pfeiffer,
Reinhard Schulte,
Manju Sudhakar
and Andrew Wroe
for helpful discussions.
M. Kuster and S. Hauf acknowledege support by the Bundesministerium 
f\"ur Wirtschaft und
Technologie and the Deutsches Zentrum f\"ur Luft- und
 Raumfahrt - DLR under the grant number 50QR0902.

\end{document}